\documentstyle[11pt]{article}

\newcommand{\eq}{\begin{equation}}
\newcommand{\en}{\end{equation}}
\newcommand{\eqn}{\begin{eqnarray}}
\newcommand{\enn}{\end{eqnarray}}

\newcommand{\beq}{\begin{equation}}
\newcommand{\eeq}{\end{equation}}
\begin{document}
\begin{titlepage}
\begin{flushright}
  PSU-TH-194\\
\end{flushright}
\begin{center}
{\bf Singletons, Doubletons  and  
M-theory} \\
\vspace{1cm} 
{\bf M. G\"{u}naydin\footnote{Work supported in part by the
National Science Foundation under Grant Number PHY-9631332. \newline 
e-mail: murat@phys.psu.edu}
 and D. Minic} \footnote{ 
e-mail: minic@phys.psu.edu}  \\
\vspace{.5cm}
Physics Department \\
Penn State University\\
University Park, PA 16802 \\
\vspace{1cm}
{\bf Abstract}
\end{center}
We identify the two dimensional AdS subsupergroup $OSp(16/2,R)$
of the M-theory supergroup $OSp(1/32,R)$ 
which captures the dynamics of $n$ $D0$-branes in the large $n$
limit of Matrix theory. The $Sp(2,R)$ factor in the  even subgroup $SO(16) \times Sp(2,R)$
of $OSp(16/2,R)$ corresponds to the 
AdS extension of the Poincare symmetry of the longitudinal
directions. The infinite number of $D0$-branes with ever increasing
and quantized values of longitudinal momenta are identified with the 
Fourier modes of the singleton supermultiplets of $OSp(16/2,R)$,
which consist of 128 bosons and 128 fermions. The large $n$
limit of $N=16$ $U(n)$ Yang-Mills quantum mechanics which
describes Matrix theory is a conformally invariant $N=16$
singleton quantum mechanics living on the boundary of $AdS_{2}$.
We also review some of the earlier results on the spectra of Kaluza-Klein
supergravity theories in relation to the recent conjecture of Maldacena
relating the dynamics of  $n$ $Dp$-branes to certain AdS supergravity
theories. We point out the remarkable parallel between the conjecture of Maldacena 
and the construction of the spectra of $11-d$ and type $IIB$ supergravity theories
compactified over various spheres in terms of singleton or doubleton supermultiplets
of corresponding AdS supergroups.

\end{titlepage}

\renewcommand{\theequation}{\arabic{section} - \arabic{equation}}
\section{Introduction}
\setcounter{equation}{0}

In the late twenties and early thirties the
works of Dirac, Heisenberg, Jordan, Pauli and
other physicists lay the foundations of relativistic 
quantum field theory. 
Feynman, Schwinger, Tomonaga and Dyson  
developed the successful theory of quantum electrodynamics
in the late forties.
In 1954 Yang and Mills formulated
non-Abelian gauge theories and  in the early seventies 
't Hooft  proved the renormalizability
of spontaneously broken non-Abelian gauge theories. 
The successful standard
 model of elementary
particles is  a local relativistic spontaneously broken non-Abelian gauge
theory in which
local fields transform covariantly under the Lorentz
group. 

About a decade after the foundations of 
relativistic quantum field theories were laid
 Wigner constructed and classified all unitary
irreducible
representations (UIR) of the Poincare group \cite{wigner}.
In Wigner's point of view, particles are represented by states in a 
positive definite Hilbert space which carries  UIRs of the Poincare group.
The transition from Wigner's representation
 of the Poincare group to the relativistic fields transforming
in the finite dimensional non-unitary representations of the Lorentz
group is  achieved by Fourier expanding the covariant fields in
terms of particle annihilation and creation 
operators that operate in a positive
definite Hilbert space \cite{sw}. The particle
states carry the unitary representations of the Poincare group.

Similarly, in a spacetime with constant curvature
(de Sitter and anti-de Sitter (AdS) space-times) 
\cite{hawk} the particle states form 
the basis of a unitary representation of
de Sitter and AdS groups. It is well known that
de Sitter groups $SO(d,1)$ for $d >2$ in d-dimensional space do not
admit unitary irreducible representation of the lowest
(or highest) weight type . 
On the other hand AdS groups $SO(d-1,2)$ do admit UIR of the
highest (or lowest weight) type  
(see \cite{mgcs} and references therein). The  operator whose
spectrum is bounded from below  in a lowest  weight 
representation  is the energy operator.
In general, local spacetime supersymmetry
in a space of constant curvature (for $d>2$)
requires that the spacetime be of AdS type \cite{ss}. 
In a relativistic quantum field theory in an AdS background the fields
transforming covariantly under the Lorentz group can be expanded
in terms of particle creation and annihilation operators.
The states created by the particle creation operators carry unitary
representation of the AdS group $SO(d,2)$.

Simple AdS supergroups exist in $d$ dimensional space-times with 
$d \le 7$  \cite{wn}.  In $d=3$ the AdS group is not simple
and decomposes as $SO(2,2) \approx SO(2,1)_{+} \times SO(2,1)_{-}$.
As a consequence the AdS group  in three 
dimensions admits $(p,q)$ type supersymmetric extensions which correspond to
$p$ and $q$ supersymmetric extensions of $SO(2,1)_{+}$ and 
$SO(2,1)_{-}$, respectively.
A complete list of AdS supergroups in $d=3$  was given in  \cite{gst}.

Below we list the AdS supergroups in $d\le 7$ and their even
subgroups:

$d=7 \qquad OSp(8^{*}/2N) \supset SO^{*}(8) \times USp(2N)$

$d=6 \qquad F(4) \supset SO(5,2) \times SU(2)$

$d=5 \qquad SU(2,2/N) \supset SU(2,2) \times U(N)$

$d=4 \qquad OSp(N/4,R) \supset SO(N) \times Sp(4,R)$

$d=3 \qquad G_{+} \times G_{-}$,

$d=2 \qquad G_{+} $,

where $G_{+}$ and $G_{-}$ can be any one of the following supergroups:

i) $ OSp(N/2,R) \supset O(N) \times Sp(2,R)$

ii) $SU(N/1,1) \supset U(N) \times SU(1,1), N \neq 2 $

\quad $SU(2/1,1) \supset SU(2) \times SU(1,1)$ 

iii) $OSp(4^{*}/2N) \supset O^{*}(4) \times USp(2N) = SU(2) 
\times USp(2N) \times SU(1,1)$

iv) $G(3) \supset G_{2} \times SU(1,1)$, with $G_{2}$ compact

v) $F(4) \supset Spin(7) \times SU(1,1)$ with $Spin(7)$ compact

vi) $D^{1}(2,1,\alpha) \supset SU(2) \times SU(2) \times SU(1,1)$

Some of the above AdS supergroups arise 
as the symmetry supergroups of  ground states of gauged supergravity 
theories in the respective
dimensions as well as the Kaluza-Klein compactifications 
on  manifolds with  non-trivial isometry
groups. For example, the ground state of the $S^{7}$ (or $S^{4}$)
compactification of the eleven-dimensional supergravity is
$AdS_{4} \times S^{7}$ (or $AdS_{7} \times S^{4}$).
Their respective symmetries are $OSp(8/4,R)$ 
\cite{dfhn,mgnw} and $OSp(8^{*}/4)$
\cite{gnw}.
Similarly II B supergravity in ten dimensions
admits compactification down
to five dimensions on $S^{5}$. 
The symmetry of the corresponding ground state is
$SU(2,2/4)$ \cite{mgnm}.

\section{ Singleton and Doubleton Supermultiplets of AdS Supergroups in 
Supergravity  and Superstring Theories}
\setcounter{equation}{0}

In  \cite{mgcs} a general oscillator method
was developed for constructing the unitary irreducible
representations (UIR) of the
lowest ( or highest) weight type of non-compact groups. 
This method was generalized
in \cite{ibmg} to the constuction of UIRs of
non-compact supergroups. This simple oscillator method is
quite powerful and yields readily the UIRs of lowest weight type of
all non-compact groups and supergroups. Among these representations
there exist some fundamental ones, in terms of which all the other
ones can be constructed by simple tensoring procedure.
These representations can be looked upon as the "quarks"
of various non-compact groups and supergroups. In the case of
 $d=4$  AdS group
$SO(3,2)$ these  fundamental representations were discovered
by Dirac long time ago \cite{pam}, 
which he referred to as the remarkable representations 
of AdS group. These representations of Dirac
were later called singletons \cite{fron} (in the
oscillator language these representations require
a single set of oscillators transforming
in the fundamental representation of the maximal compact subgroup $SU(2) \times U(1)$ 
of the covering group 
$Sp(4,R)$ of
$SO(3,2)$ \cite{mgnw,mg81,mg}. 
The corresponding
fundamental representations
of $d=7$ and $d=5$ AdS groups $ SO^*(8)$ and $SU(2,2)$, respectively, were named doubletons in 
\cite{gnw,mgnm} since they require two 
sets of oscillators transforming
in the fundamental representations of  
corresponding maximal compact subgroups.

The oscillator method for non-compact groups $G$ that admit
lowest weight representations works as follows:
$G$ has a maximal compact subgroup $G^{0}$ which is of the form
$G^{0} = H \times U(1)$ with respect to whose Lie algebra $g^0$  we have
a three grading of the Lie algebra $g$ of $G$,
\eq
g = g^{-1} \oplus g^{0} \oplus g^{+1}
\en
which simply means that the commutator of elements of grade
$k$ and $l$ satisfies
\eq
[g^{k},g^{l}] \subseteq g^{k+l}
\en

For example, for $SU(1,1)$ this corresponds to the standard
decomposition
\eq
g = L_{+} \oplus L_{0} \oplus L_{-}
\en
where
\eqn
[L_{0},L_{\pm}] &= &\pm L_{\pm} \cr
[L_{+},L_{-}] &=&2L_{0} 
\enn

The three grading is determined by the generator $E$ of the
$U(1)$ factor of the maximal compact subgroup
\eqn
[E,g^{+}] &=&g^{+} \cr
[E,g^{-}] &=&-g^{-} \cr
[E,g^{0}] &=& 0 
\enn

In most physical applications $E$  turns out
to be the energy operator. 
In such cases the unitary lowest weight representations
correspond to positive energy representations.

The essence of the oscillator
method is to realize the generators of $G$ as
bilinears of bosonic annihilation
and creation operators transforming typically in the fundamental
(and its conjugate) representation of $H$.
In the Fock space $\cal{H}$ of all the oscillators one 
chooses a set of  states
$|\Omega \rangle$ which transform irreducibly under $H \times U(1)$
and are annihilated by all the generators in $g^{-1}$.
Then by acting on $|\Omega \rangle$ with generators in
$g^{+1}$ one obtains an infinite set of states
\eq
|\Omega \rangle ,\quad  g^{+1}|\Omega \rangle ,\quad
g^{+1} g^{+1}|\Omega \rangle , ...
\en
which form the UIR of the lowest weight type (positive energy )
of $G$. The infinite set of states thus obtained corresponds to the 
decomposition of the UIR of $G$ with respect to its maximal
compact subgroup.

As we have already mentioned, whenever
 we can realize the generators of $G$ in
terms of a single set of oscillators 
transforming in an irreducible representation of 
$H$ the corresponding
UIRs of the lowest weight type will be called singleton representations and 
 there exist two such representations for a given group 
 $G$. For AdS groups the singleton
representations correspond to scalar and spinor fields .
In certain cases  we need  a minimum of two sets of oscillators 
transforming irreducibly under $H$ to realize the
generators of $G$. In such cases  the 
corresponding UIRs  are called doubleton
representations and there exist 
infinitely many doubleton representations 
of $G$ corresponding to fields of different 
"spins". 
For example, the non-compact
group $Sp(2N,R)$ admits singleton representations.
On the other hand, non-compact groups $SO^{*}(2N)$ and
$SU(N,M)$ admit doubleton representations. 

The oscillator method can  similarly be used to construct unitary lowest
type UIR's of non-compact
supergroups \cite{ibmg,mg88}. They also admit singleton or doubleton
supermultiplets corresponding to some fundamental 
lowest weight unitary irreducible
representations.
For example, the non-compact supergroup $OSp(2N/2M,R)$ with the even subgroup
$SO(2N)\times Sp(2M,R)$  admits singleton
supermultiplets, while $OSp(2N^*/2M)$ and $SU(N,M/P)$ with even subgroups 
$SO^*(2N) \times USp(2M)$ and $SU(N,M)\times SU(P) \times U(1)$ admit
doubleton supermultiplets.

As is well known the AdS group $SO(d-1,2)$ admits an In\"{o}n\"{u}-Wigner
type contraction to the Poincare group in $d$-dimensional Minkowskian
space-time limit.
What is quite  surprising is the fact that the singleton
representations of Dirac in $d=4$ are singular when the Poincare
limit is taken!
This is due to the remarkable property noticed by
Dirac \cite{pam}, that the wave functions
corresponding to the singleton representations do not
depend on the radius of the AdS space-time. Singletons can
be thus properly understood as fields living on the
boundary of the corresponding AdS space-time. Yet the 
boundary singleton field theory controls the physics of
the "bulk" AdS space-time!.
The AdS group $SO(3,2)$ should
be interpreted as the conformal group of the singleton
theory living on the
boundary of the AdS space-time.

That the "remarkable" representations of Dirac \cite{pam}
do not have a Poincare
limit has been known for a long time \cite{fron}. One way to see
that singleton irreducible representations can not have a 
Poincare limit is to look at the singleton supermultiplets.
For example, the singleton supermultiplet of $OSp(8/4,R)$ has
8 scalar and 8 spin ${1 \over 2}$ fields \cite{mgnw,mg,mg81}. On the
other hand, it is well known 
that the shortest $N=8$ Poincare supermultiplets have spin range two.
Similarly one can show that singleton or doubleton supermultiplets
of extended AdS supergroups in various dimensions can not have a
Poincare limit.
 
The unitary supermultiplets of AdS supergroups in $d=3$ were
constructed in \cite{gst} and in \cite{gnst} it was shown 
that the light-cone Green-Schwarz
II A, II B and heterotic superstring theories can be interpreted as the
(${\bf 8_{c}, 8_{s}}$), (${\bf 8_{c}, 8_{c}}$) and (${\bf 8_{c}, 0}$)
superconformally invariant singleton field theories of the
$AdS_{3}$ supergroups $OSp(8/2,R)_{c} \times OSp(8/2,R)_{s}$,
$OSp(8/2,R)_{c} \times OSp(8/2,R)_{c}$ and
$OSp(8/2,R)_{c} \times Sp(2,R)$, respectively.

In \cite{mgnw} the unitary 
supermultiplets of $N=8$ AdS supergroup $OSp(N/4,R)$
in $d=4$  
were constructed and the spectrum of the $S^{7}$ compactification
of  eleven dimensional supergravity was fitted into  its unitary
supermultiplets. The singleton supermultiplet sits
at the bottom of an infinite tower of Kaluza-Klein modes and 
appears as gauge degrees of freedom that decouple from the spectrum 
\cite{mgnw}. However, in 
a related work it was shown that the infinite tower of Kaluza-Klein
states of any given spin fall into  irreducible representations
of a spectrum generating non-compact group $SO(8,1)$ \cite{grw} .
This infinite tower of Kaluza-Klein states contains all the physical
modes as well as gauge modes including the singleton modes.
Thus singleton modes are essential for fitting the Kaluza-Klein 
states into unitary representations of a spectrum generating
group.  Furthermore, even though the singleton supermultiplet 
decouples from the spectrum as gauge modes, by tensoring the singleton
supermultiplets repeatedly and restricting oneself to CPT self-conjugate 
supermultiplets in the tensor products one generates the entire tower
of Kaluza-Klein supermultiplets of 11 dimensional supergravity 
over the seven sphere!

The compactification of 11-d supergravity over the four 
sphere $S^{4}$ down to seven dimensions was studied in 
\cite{gnw,ptn}.
The ground state of this compactification is 
$AdS_{7} \times S^{4}$ and has the symmetry $OSp(8^{*}/4)$
with the even subgroup $SO(6,2) \times USp(4)$.
The infinite Kaluza Klein modes fall into unitary supermultiplets
of $OSp(8^{*}/4)$ \cite{gnw} .
Again the doubleton supermultiplet appears as 
gauge modes and decouples from the  
spectrum. It consists of five scalars, four fermions and
a self-dual two form field. 
The infinite set of Kaluza-Klein modes of a given spin together with
some gauge modes form the basis of a unitary representation of
a spectrum generating group $SO(5,1)$  \cite{grw,mgunp}.
The doubleton supermultiplet sits at the bottom of the infinite tower of
$SO(5,1)$ modes  corresponding to scalar, spinor and the
self-dual antisymmetric tensor fields .

The unitary supermultiplets of  $2N$ extended  
$AdS_{5}$ supergroups $SU(2,2/N)$
were studied in \cite{mgnm,ibmg}.  
The spectrum of the $S^{5}$ compactification of ten dimensional
IIB supergravity was calculated in \cite{mgnm,krv}. 
The entire spectrum  
falls into an infinite tower of unitary supermultiplets of
$N=8$ $AdS_{5}$ superalgebra $SU(2,2/4)$ \cite{mgnm}.
The $CPT$ self-conjugate doubleton supermultiplet of 
$N=8$ $AdS$ superalgebra again decouples from the physical 
spectrum and corresponds to some gauge modes.
In \cite{mgnm} it was pointed out that $N=8$ 
doubleton field theory is the conformally invariant 
$N=4$ super Yang-Mills theory in $d=4$.
Again, even though the doubleton supermultiplet
decouples from the Kaluza-Klein spectrum of ten
dimensional II B supergravity on
$S^{5}$, by tensoring the doubleton supermultiplet with itself 
repeatedly and restricting oneself to the $CPT$ self-conjugate
supermultiplets one generates the entire spectrum of
Kaluza-Klein states of ten dimensional II B theory on $S^{5}$. 
As in the case of  $S^{7}$ and $S^{4}$ compactifications of
11-d supergravity one can show that the Kaluza Klein modes of a 
given spin together with some gauge modes  form unitary 
representations of  a spectrum generating $SO(6,1)$ 
in the $S^{5}$ compactification of
ten dimensional IIB supergravity \cite{grw,mgunp}.

Even though the Poincare limit of 
the singleton representations of the $d=4\  AdS$ group
$SO(3,2)$ is singular the tensor product of two singleton irreps of   
$SO(3,2)$ decomposes into an infinite set of 
massless irreps which do have a    
smooth Poincare limit \cite{fron,mg81,mgnw} . 
Similarly, the tensor product of two            
singleton supermultiplets of $N$ extended $AdS$ supergroup $OSp(N/4,R)$         
decomposes                                                                      
into an infinite set of massless supermultiplets which do have a Poincare   
limit \cite{mgnw,mg81} . The simple $AdS$ groups in 
higher dimensions than four that do admit    
supersymmetric extensions have doubleton representations only. The doubleton    
supermultiplets of extended $AdS$ supergroups in $d=5\ (SU(2,2/N))$ and         
$d=7\ (OSp(8^*/2N))$ share the remarkable features of the singleton             
supermultiplets of $d=4$ $AdS$ supergroups 
in that the tensor product of any two       
doubleton                                                                       
supermultiplets decompose into an infinite set of massless supermultiplets      
\cite{gnw,mgnm}.  Based on these facts and some other arguments one 
of the authors made
the following proposal for defining massless 
representations \cite{mg} in AdS
spacetimes:
                                                                         
{\bf A representation (or a supermultiplet) 
of an $AdS$ group (or supergroup) is
 massless if it occurs in the decomposition of 
the tensor product of  
two singleton or two doubleton representations (or supermultiplets).}           

This should be taken as a working definition and agrees with some other definitions
of masslessness in $d \leq 7$.  Tensoring more than two singleton or doubleton representations 
leads to massive
representations  of AdS groups and supergroups \cite{mgnw,gnw,mgnm,mg}.

\section{ Singletons, Doubletons and p-branes}
\setcounter{equation}{0}

Gibbons and Townsend \cite{gt} found solutions of
$d=10$ and $d=11$ supergravity equations of
motion that interpolate between 
different supersymmetric
vacua connected via a wormhole throat. In particular,
in $d=11$ the relevant p-branes are membrane and five-brane and
in $d=10$ II B self-dual three-brane configuration. 
These p-branes interpolate between d-dimensional Minkowski
space-time  and $AdS_{p+2} \times S^{d-p-2}$ space. The p-brane
solutions are non-singular and break half of the supersymmetries.

Now, $AdS_{p+2} \times S^{d-p-2}$ are, as pointed out
 above, known to be maximally supersymmetric solutions of the 
corresponding $d=10$ and $d=11$ supergravity theories.
Furthermore, effective world-volume actions for these p-brane solutions in
$d=10$ and $d=11$ are related to singleton supermultiplets of the
corresponding $AdS$ supergroups. In particular, the fields of the
various singleton supermultiplets, already listed above, show up
as corresponding factors in the harmonic expansion of the d-dimensional
fields on the $d-p-2$ spheres. As we have pointed out, these modes
can be gauged away everywhere except on the boundary of the
appropriate $AdS_{p+2}$ space-time. These boundaries represent
nothing more than p-brane cores, or opening of the 
wormhole throat that connects d-dimensional Minkowski space-time
with the corresponding $AdS$ space-time \cite{gt}. The relevant singleton
supermultiplets, as we have seen, are in 3-d the $N=8$ scalar
supermultiplet (for $AdS_{4}$) \cite{mgnw,nst,bsst},
 in 6-d the $N=2$ antisymmetric tensor 
supermultiplet (for $AdS_{7}$) \cite{gnw}, in 4-d the $N=4$ Yang-Mills 
supermultiplet (for $AdS_{5}$) \cite{mgnm,sfcf}. 
The first two arise in the
analysis of the quadratic fluctuations around eleven dimensional
membrane  \cite{nst,bsst,bdps,dwhn,bst},
and five-brane solutions 
\cite{fiveb,calt,ckvp}.
The corresponding singleton superconformal theory controls the
physical excitations of the eleven dimensional p-brane world-
volume theories. 

Based on these observations and closely related results of 
\cite{gibb,mbmd,asjm,hyun}, 
Maldacena \cite{mald}
considered dynamics
of $n$ parallel $Dp$-branes in the limit when the field theory on
the $Dp$-brane decouples from the bulk. 
 Maldacena argued that in
the limit of
the near horizon geometry supergravity solutions pointed out
by Gibbons and Townsend ($AdS_{p+2} \times
S^{d-p-2}$) can be trusted if the large $n$ limit
is taken. The curvatures of $AdS_{p+2}$ space and $S^{d-p-2}$
scale as positive powers of $1/n$, so in the large $n$ limit 
these supergravity solutions are valid. The Hilbert space of the
induced (singleton or doubleton) 
superconformal field theories includes the Hilbert
space of supergravity solutions of $AdS_{p+2} \times S^{d-p-2}$.
Based on this observation Maldacena then conjectured that
compactifications of M-theory (or superstring theory) on 
$AdS_{p+2} \times S^{d-p-2}$ are dual to various 
superconformal field theories. (These observations were
extended in a number of recent works
 \cite{sfcf,kkr,andy,kb,skend,kall}.) 

Our analysis above show that these superconformal theories are
simply the singleton or doubleton field theories of AdS supergroups
in various dimensions.
Furthermore we would like to point out that the conjecture of Maldacena has 
a beautiful counterpart in the construction of massless and massive
irreducible $N=8$ $AdS_{5}$ supermultiplets in terms of doubleton
supermultiplets. If we make the assumption that the bound 
states of the doubleton field theory i.e. the $N=4$ Yang-Mills
in $d=4$ (boundary of $AdS_{5}$) are CPT self-conjugate
supermultiplets then the two doubleton bound states give 
the massless $N=8$ $AdS_{5}$ 
supergravity multiplet. These bound states have twice the $AdS_{5}$ 
energy of doubletons and correspond to massless particles in
the Poincare limit! Bound states of three and more 
doubletons yield ever increasing tower of massive supermultiplets
which correspond to the spectrum of 
ten dimensional IIB supergravity over $S^{5}$ \cite{mgnm}.
What is remarkable is that these bound states of doubletons
are bound states at threshold i.e. with no binding energy
(in the $AdS$ sense!!). This is exactly the picture
proposed by Witten for the bound states of $Dp$-branes $\cite{witten}$.
Of course, as is well known, in ordinary QCD we do not expect to have 
bound states at threshold. The above picture suggests 
that the $N=4$  $U(n)$ Yang-Mills theory
must have bound states at threshold as $n$ goes to infinity.

This picture also holds true for the doubleton field 
theory of $N=4$ $AdS_{7}$ supermultiplet in $d=6$. The 
massless and massive bound states (again assuming  CPT invariance as above)
 yield the complete
spectrum of eleven- dimensional supergravity compactified 
down to $d=7$ over $S^{4}$.
Similarly, the singleton field theory of $N=8$
$AdS_{4}$ bound states in $d=3$ yields the spectrum
of eleven dimensional supergravity over $S^{7}$.
In all these cases, the $AdS$ bound states are bound states
at threshold in the sense of AdS energy!

\section{Singletons, Doubletons and Matrix Theory}
\setcounter{equation}{0}

M-theory \cite{mth} has as its low energy theory the eleven
dimensional supergravity. It includes in its spectrum the
states of eleven dimensional supergravity and BPS
states that correspond to the extension of eleven dimensional
SUSY algebra by central charges.
The eleven dimensional M-theory SUSY algebra with two-form and
five-form central charges can be obtained as a 
contraction of the simple supergroup $OSp(1/32,R)$
\cite{vhvp}, \cite{town}, with the
even subgroup $Sp(32,R)$. It can be written in the form 
\eq
\{Q_{\alpha},Q_{\beta}\} = (C\Gamma^{m})_{\alpha \beta} P_{m}+ 
{1 \over 2}(C\Gamma^{m_1 m_2})_{\alpha \beta}Z_{m_1 m_2}  +
{1 \over {5!}}(C \Gamma^{m_1 ... m_5})_{\alpha \beta}Z_{m_1...m_5} 
\en
where $m_{i} = 0,...,10$ and $\alpha, \beta =1,...,32$; $\Gamma^{m}$
are Dirac matrices, $\Gamma^{m_1...m_k} (k=2,5)$ are their antisymmetrized
products and $C$ is the charge conjugation matrix.

Matrix theory as proposed by Banks, Fischler, Shenker and
Susskind \cite{bfss} corresponds to a formulation of M-theory in the
infinite momentum frame (IMF) .
In the IMF the Poincare symmetry is broken down to
Galilean symmetry and the SUSY algebra takes the
form   
 \cite{tbss} 
\eqn
\{Q^{+}_{\mu},Q^{+}_{\nu} \} &=&  P^{+} \delta_{\mu \nu} \cr 
\{Q^{-}_{\mu},Q^{-}_{\nu}\}  &=& P^{-} \delta_{\mu \nu}
+ \Gamma_{\mu \nu}^{a} Z_{a} + 
{1 \over {4!}} \Gamma^{a_1...a_4}_{\mu \nu} Z_{a_1...a_4} \cr
\{Q^{+}_{\mu},Q^{-}_{\nu}\} &=& \Gamma_{\mu \nu}^{a} P_{a} 
+ {1 \over 2} \Gamma^{a_1 a_2}_{\mu \nu} Z_{a_1 a_2} + 
 {1 \over {5!}} \Gamma^{a_1...a_5}_{\mu \nu}Z_{a_1...a_5}
\enn
where the original thirty two supercharges split into a
pair of sixteen supercharges $Q^{+}_{\mu}$ and $Q^{-}_{\mu}$
($\mu=1,...,16$) and $a_i$  are the transverse space indices. 

In the IMF limit, as pointed out by Banks, Fischler, Shenker and
Susskind, all mass scales
disappear and the spectrum consist of $D0$-branes \cite{joe} which
correspond to plane eleven dimensional gravitational waves.
They are described by the translation part of the algebra above.

The dynamics of $n$ $D0$-branes on the other hand is described by $N=16$ SUSY
$U(n)$ quantum mechanics corresponding to 128 bosons and
128 fermions (called supergravitons in \cite{bfss}) 
\cite{halp,witten,dkps} with the Lagrangian ${\cal{L}} =Tr L$ where 
\eq
L = {1 \over {2R}}(D_0 X^a)^2 + \theta^{\alpha} D_0 \theta_{\alpha}
-{R \over 4} {[X^a,X^b]}^{2} + iR  \theta^{\alpha}
{\Gamma_{a \alpha \beta}} [X^a, \theta^{\beta}] 
\en
where $R$ is the extent of the
longitudinal direction,  
$D_0 \equiv \partial_0 - [A_0,]$ and $A_0, X^a (a=1,...,9)$,
$\theta^{\alpha}$
$(\alpha = 1,...16)$ are hermitean $n \times n$ matrices. The
longitudinal momentum is $P^{+} = n/R$. 

Let us  now identify an $OSp(16/2,R)$  subsupergroup 
 of $OSp(1/32,R)$ that will be important 
to the understanding of the dynamics
of $D0$ particles in the IMF.
The Lie superalgebra of $OSp(1/32,R)$ 
can be given a 5-graded structure with
respect to its maximal compact subalgebra $U(16)$ \cite{mg88}
\eqn 
OSp(1/32,R) &=& L_{ij} \oplus L_i \oplus L^{i}_{j} \oplus L^i \oplus L^{ij} \cr
          &=& g^{-2} \oplus g^{-1} \oplus g^0 \oplus g^1 \oplus g^2
\enn
where $L^{i}_{j}$ are $U(16)$ generators and $L_i (L^i)$ transform
in the fundamental $16 (\bar{16})$ representation of $U(16)$. The generators
$L_{ij} = L_{ji}$ transform in the symmetric tensor representation 
of $U(16)$. We also have $L_{ij}^{\dagger}= L^{ij}$ and
$L_i^{\dagger}=L^i$.

The trace components of the grade $\mp 2$  and grade 
zero subspaces $( L_{ii}, L^{ii}, L^i_i)$
form an $Sp(2,R)$ subalgebra. The maximal subalgebra of $U(16)$
that commutes with this $Sp(2,R)$ is $SO(16)$.
The $SO(16) \times Sp(2,R)$ subalgebra has an extension to
the superalgebra $OSp(16/2,R)$ which is a subalgebra of $OSp(1/32,R)$.
We shall identify this $Sp(2,R)$ with the $AdS$ extension of 
the Poincar\'{e} group of the light-cone coordinates.
With this identification $OSp(16/2,R)$ becomes the $AdS$ symmetry of 
the  M-theory algebra when it is compactified down to two
dimensions. 

The supergroup $OSp(16/2,R)$ admits singleton supermultiplets \cite{gst}.
The corresponding field theory exists only on the boundary
of the two dimensional $AdS_{2}$ space and 
is simply the conformally invariant singleton field theory
in one dimension (more precisely on the boundary
of $AdS_{2}$) with 16 supercharges. 
The singleton supermultiplets of $OSp(16/2,R)$ consists of 
 $128$ bosons and $128'$ fermions, 
transforming in the left-handed ($128$) and right-handed
($128'$) spinor representation of $SO(16)$. In the other
singleton supermultiplet the bosons transform in $128'$ 
and fermions in $128$ of $SO(16)$. 

Singleton representation of $Sp(2,R)$ when expanded in a 
particle basis consists of an infinite tower of states with an ever 
increasing quantized $U(1)$ eigenvalues \cite{gst}.
We propose to identify this infinite tower of states 
with an infinite tower of $D0$-branes \cite{joe} with quantized 
values of the longitudinal momentum as in the Matrix theory 
framework of
M-theory in the infinite momentum frame  \cite{bfss} i.e.
the singleton supermultiplets of $OSp(16/2,R)$ can be 
identified with an infinite tower of $128 + 128$ $D0$-branes 
(supergravitons).  

Note that to get a precise correspondence with singleton 
field theory we need an infinite number of supergravitons.
This is also consistent with the claim that for large $n$ 
the dynamics of $D0$-branes is related to the light-cone 
supermembrane \cite{dwhn}.
More
specifically, Matrix theory of $N=16$ Yang-Mills quantum
mechanics in the limit of the infinite number ($n=\infty$)
of $D0$-branes is given by the $2+1$ dimensional light-cone
Hamiltonian for the classical supermembrane configuration.
The Lorentz symmetry of the corresponding world-volume
theory of the supermembrane \cite{bst} can then be identified with our
$Sp(2,R) \approx SO(2,1)$ symmetry.
Thus $OSp(16/2,R)$ conformal singleton field theories represent the
large $n$ limit of Matrix theory. 

This picture is in perfect agreement with Maldacena's proposal  
\cite{mald} for the case of the large $n$ number of
$D0$-branes. Thus the singleton quantum mechanics 
living on the boundary of $AdS_{2}$ controls the
large $n$ limit of Matrix theory  which is 
the $N=16$ $U(n)$ Yang-Mills quantum mechanics. 
Furthermore this  is a rather natural
extrapolation of the correspondence
between $AdS_{3}$ singleton field theories 
and string theories (IIA, IIB and
heterotic) in the light-cone frame \cite{gnst}. 

The fact that the singleton quantum mechanics is conformally invariant
is consistent with the requirement that in the large $n$ limit of Matrix theory
longitudinal boost invariance (which is closely related
to scale invariance in the longitudinal momentum) has to be recovered,
thus restoring the full 11-d Lorentz invariance.

\section{Discussion and Conclusions}
\setcounter{equation}{0}

Interpreting the simple supergroup $OSp(1/32,R)$ as a
"generalized $AdS$" supergroup in $d=11$ \cite{fre} whose contraction
gives the M-theory superalgebra \cite{vhvp,town}, suggests 
immediately that it can also be understood as the 
generalized conformal group in ten dimensions \cite{town}.
Now, the noncompact supergroups $OSp(2N+1/2M,R)$ with the 
even subgroup $SO(2N+1) \times Sp(2M,R)$ do not in general
admit three grading with respect to a compact subsupergroup
. However $OSp(2N+1/2M,R)$ does admit a five-grading
with respect to its subsupergroup $U(N|M)$ \cite{mg88}. The oscillator
method of \cite{ibmg} for constructing UIR's of non-compact
supergroups was generalized to supergroups that admit
such five-grading in \cite{mg88}. Using this method one can
construct the unitary supermultiplets of $OSp(1/32,R)$ \cite{mg98}.
Positive energy UIR's of $OSp(1/32,R)$ are uniquely
characterized by the $U(16)$ quantum numbers of the 
the corresponding lowest weight vectors. One finds that 
$OSp(1/32,R)$ admits a singleton supermultiplet consisting of a 
scalar field $\phi(1)$ and a spinor field $\psi(16)$ \footnote{we
indicate the $SU(16)$ representation of the fields in
parenthesis while omitting their $U(1)$ charges.}
 
Some of the lowest "massless" supermultiplets of
$OSp(1/32,R)$ are \cite{mg98} \\

$ \phi(1) \oplus \psi_{\mu}(16) \oplus A_{\mu \nu}(120) $ \\

$ \psi_{\mu}(16) \oplus S_{\mu \nu}(136)   $  \\

$ S_{\mu \nu}(136) \oplus \psi_{\mu \nu \rho}(816) $ \\

$ \psi_{\mu \nu \rho}(816) \oplus S_{\mu \nu \rho \lambda}(3876)  $ \\

where $\mu, \nu,...=1,2,...,16$ are $SU(16)$ indices.  

To get a supermultiplet of $OSp(1/32,R)$ which includes the fields of 
11-d supergravity one needs to tensor four copies of the 
singleton supermultiplet \cite{mg98}.
However, following the definition of "masslessness" stated
in section two \cite{mg} these would correspond to massive
supermultiplets! 
One possible resolution of this problem is that these "massive"
supermultiplets become massless supermultiplets of 11-d
Ponicare superalgebra when we contract $OSp(1/32,R)$.
Furthermore, the definition of masslessness proposed in
\cite{mg} may have to be modified in the presence of
central charges. That the contraction of $OSp(1/32,R)$ to
11-d supergravity algebra may change the definition of
masslessness is also consistent with the fact that there
are no known 11-d supersymmetric field theories that correspond
to "massless" supermultiplets above. Another possible resolution
is to extend $OSp(1/32,R)$ to a larger symmetry such as $OSp(1/32,R) \times OSp(1/32,R)$
as suggested in some recent work \cite{horava} \cite{bars}. 
These issues will be addressed elsewhere \cite{mg98}.
 
In conclusion, in this article we have studied the singleton and doubleton
supermultiplets of AdS supergroups in various dimensions in 
relation to the dynamics of Dp-branes and Matrix theory.
We have identified the two dimensional AdS supergroup $OSp(16/2,R)$,
with the even subgroup $SO(16) \times Sp(2,R)$, which determines
the dynamics of $n$ D0-branes in the large $n$ limit of
Matrix theory. The
$Sp(2,R)$ symmetry corresponds to the AdS extension of the 
Poincare symmetry of the longitudinal direction. The
infinite number of D0-branes are  the Fourier modes of the 
singleton supermultiplets of $OSp(16/2,R)$ which consist of
128 bosons and 128 fermions. We have also shown that the recent
conjecture of Maldacena relating the dynamics of $n$ Dp-branes
to AdS supergravity theories has a beautiful counterpart in the
known constructions of the spectra of 10-d IIB and 11-d supergravity
theories over various spheres in terms of singleton and doubleton
supermultiplets. In all these cases the field theories of singleton
and doubleton supermultiplets of AdS supergroups in various dimensions
are conformally invariant theories living on the boundary of the
AdS space-time. In particular, the large $n$ limit of $N=16$ 
$U(n)$ Yang-Mills quantum mechanics which describes Matrix theory
is a conformally invariant $N=16$ singleton quantum mechanics.


\vskip.1in

\begin{thebibliography}{99}

\bibitem{wigner} E. Wigner, Ann. Math, 40 (1939) 149.
\bibitem{sw} S. Weinberg, {\it The Quantum Theory
of Fields}, vol. I and II, Cambridge University Press, 1995-96.
F. G\"{u}rsey, Theory of Quantized Fields, Yale University lecture
notes (1967) .
\bibitem{hawk}  S. W. Hawking and G. F. R. Ellis, {\it The Large Scale
Structure of Space-Time}, Cambridge (1973); S. Weinberg, {\it Gravitation
and Cosmology}, J. Wiley  (1972).
\bibitem{mgcs} M. G\"{u}naydin and C. Saclioglu, Comm. Math. Phys.
87 (1982) 159.
\bibitem{ss} A. Salam and E. Sezgin, {\it Supergravity in Different
Dimensions}, World Scientific (1989).
\bibitem{wn} W. Nahm. Nucl. Phys. B135 (1978) 149.
\bibitem{gst} M. G\"{u}naydin, G. Sierra and P. K. Townsend,
Nucl. Phys. B274 (1986) 429.
\bibitem{dfhn} D. Freedman and H. Nicolai, Nucl. Phys. B237 (1984) 342;
P. Breitenlohne and D. Z. Freedman, Ann. Phys. 144 (1982) 249;
W. Heidenreich, Phys. Lett. 110B (1982) 461.
\bibitem{mgnw} M. G\"{u}naydin and N. P. Warner, Nucl. Phys. B272
(1986) 99.
\bibitem{gnw} M. G\"{u}naydin, P. van Nieuwenhuizen and N. P. Warner,
Nucl. phys. B255 (1985) 63.  
\bibitem{mgnm} M. G\"{u}naydin and N. Marcus, Class. and Quantum
Gravity, 2 (1985) L11; {\it ibid} L19. 
\bibitem{ibmg} I. Bars and M. G\"{u}naydin, Comm. Math. Phys.
91 (1983) 21.
\bibitem{pam} P. A. M. Dirac, J. Math. Phys. 4 (1963) 901.
\bibitem{fron} C. Fronsdal,
Phys. Rev. D26 (1982) 1988; M. Flato and C. Fronsdal, J. Math. Phys.
22 (1981) 1100; E. Angelopoulos, M. Flato, C. Fronsdal and
D. Steinheimer, Phys. Rev. D23 (1981) 1278.
\bibitem{mg81} M. G\"{u}naydin, {\it Oscillator-like Unitary
Representations of Non-compact Groups and Supergroups and Extended
Supergravity Theories}, expanded version of the invited talk given
in {\it Group Theoretical Methods in Physics}, Istanbul, 1982, Ecole
Normale Superieure preprint LPTENS 83/5 and in Lecture Notes in Physics 
Vol. 180  (1983), ed by E. In\"{o}n\"{u} and M. Serdaroglu.
\bibitem{mg} M. G\"{u}naydin, in Proceedings of the Trieste
conference {\it "Supermembranes and Physics in 2+1 dimensions"},
eds. M. J. Duff, C. N. Pope and E. Sezgin, World Scientific,
1990, pp.442.
\bibitem{mg88} M. G\"{u}naydin, J. Math. Phys. 29 (1988) 1275. 
\bibitem{gnst} M. G\"{u}naydin, B. E. W. Nilsson, G. Sierra and
P. K. Townsend, Phys. Lett. B176 (1986) 45.
\bibitem{grw} M. G\"{u}naydin, L. Romans and N. Warner, Phys. Lett.
146B (1984) 401.
\bibitem{mgunp} M. G\"{u}naydin, unpublished.
\bibitem{ptn} K. Pilch, P. K. Townsend and P. van Nieuwenhuizen,
Nucl. Phys. B242 (1984) 377.
\bibitem{krv} H. J. Kim, L. J. Romans and P. van Nieuwenhuizen.
Phys. Rev. D32 (1985) 389.
\bibitem{gt} G. W. Gibbons and P. K. Townsend, Phys. Rev.
Lett. 71 (1993) 3754.
\bibitem{nst} H. Nicolai, E. Sezgin and Y. Tanii, Nucl. Phys.
B305 (1988) 483; H. Nicolai and E. Sezgin, Phys. Lett. 143B (1984)
389. 
\bibitem{sfcf} S. Ferrara and C. Fronsdal, hep-th/9712239.
\bibitem{bsst} E. Bergshoeff, A. Salam, E. Sezgin and Y. Tanii,
Nucl. Phys. B305 (1988) 497.
\bibitem{bdps} E. Bergshoeff, M. J. Duff, C. N. Pope and
E. Sezgin, Phys. Lett 199B (1987) 69. 
\bibitem{dwhn} B. de Wit, J. Hoppe and H. Nicolai, Nucl. Phys.
B305 (1988) 545.
\bibitem{bst} E. Bergshoeff, E. Sezgin and P. K. Townsend, Phys. Lett. 
189B (1987) 75; Ann. Phys. 185 (1988) 330.  
\bibitem{fiveb} I. Bandos, K. Lechner, A. Nurmagambetov, P. Pasti,
D. Sorokin and M. Tonin, Phys. Rev. Lett. 78 (1997) 4332.
\bibitem{calt} M. Aganagic, J. Park, C. Popescu and J. H. Schwarz,
Nucl. Phys. B496 (1997) 191.
\bibitem{ckvp} P. Claus, R. Kallosh and A. van Proeyen, hep-th/9711161.
\bibitem{gibb} G. Gibbons, Nucl. Phys. B207 (1982) 337; R. Kallosh and
A. Peet, Phys. Rev. D46 (1992) 5223; S. Ferrara, G. Gibbons and 
R. Kallosh, Nucl. Phys. B500 (1997) 75.
\bibitem{mbmd} M. Blencowe and M. J. Duff, Phys. Lett. 203B (1988) 229;
Nucl. Phys. B310 (1988) 387.  
\bibitem{asjm} A. Strominger and J. Maldacena, hep-th/9710014.
\bibitem{hyun} S. Hyun, hep-th/9704005.
\bibitem{mald} J. Maldacena, hep-th/9711200.
\bibitem{kkr} R. Kallosh, J. Kunar and A. Rajaraman, hep-th/9712073.
\bibitem{andy} A. Strominger, hep-th/9712251; D. Birmingham, I. Sachs
and S. Sen, hep-th/9801019; D. Birmingham, hep-th/9801145. 
\bibitem{kb} K. Behrndt, hep-th/9801058.
\bibitem{skend} K. Sfetsos and K. Skenderis, hep-th/9711138.
 H. Jan Boonstra, B. Peeters and K. Skenderis, hep-th/9801076.
\bibitem{kall} P. Claus, R. Kallosh, J. Kumar, P. K. Townsend and
A. van Proeyen, hep-th/9801206.
\bibitem{witten} E. Witten, Nucl. Phys. B460 (1995) 335.
\bibitem{mth} C. Hull and P. K. Townsend, Nucl. Phys. B438 (1995) 109;
E. Witten, Nucl. Phys. B443 (1995) 85; for review and references consult,
for example, P. K. Townsend, hep-th/9612121.
\bibitem{vhvp} J. W. van Holten and A. van Proeyen, J. Phys.
A15 (1982) 3763.
\bibitem{town} P. K. Townsend, hep-th/9708034; D. Sorokin and P. K.
Townsend, hep-th/9709007.
\bibitem{bfss} T. Banks. W. Fischler, S. H. Shenker and
L. Susskind, Phys. Rev. D55 (1997) 5112. For reviews consult,
for example, T. Banks, hep-th/9710231; D. Bigatti and L. Susskind,
hep-th/9712072.
\bibitem{tbss} T. Banks, N. Seiberg and S. Shenker, Nucl. Phys. B490 (1997)
91.
\bibitem{joe} J. Polchinski, Phys. Rev. Lett. 75 (1995) 4724;
J. Dai, R. Leigh and J. Polchinski, Mod. Phys. Lett. (1989).
J. Polchinski, {\it TASI Lectures on D-branes}.
\bibitem{halp} M. Baake, P. Reinicke, V. Rittenberg, J. Math. Phys.
26 (1985) 1070; R. Flume, Ann. Phys. 164 (1985) 189; C. Claudson and
M. B. Halpern, Nucl. Phys. B250 (1985) 689.
\bibitem{dkps} M. R. Douglas, D. Kabat, P. Pouliot and 
S. Shenker, Nucl. Phys. B485 (1996) 85 .
\bibitem{fre} R. D'Auria and P. Fr\'{e}, Nucl. Phys. B201 (1982) 101;
L. Castellani, R. D'Auria and P. Fr\'{e}, {\it Supergravity and
Superstrings: A Geometric Perspective}, World Scientific (1991).
\bibitem{mg98} M. G\"{u}naydin, work in progress. 
\bibitem{horava} P. Horava, hep-th/9712130.
\bibitem{bars} I. Bars, Phys. Rev. D55 (1997) 2373.
\end{thebibliography}
\end{document}